\documentclass[aps,prl,twocolumn,superscriptaddress]{revtex4-1}
\usepackage[english]{babel}
\usepackage{graphicx}
\usepackage{dcolumn}
\usepackage{amsmath}
\usepackage{epsfig}
\usepackage{bm}
\usepackage{amssymb}
\usepackage{float}
\usepackage{natbib}
\usepackage{color}
\usepackage{soul}
\usepackage{hyperref}
\hypersetup{colorlinks=true,linkcolor=blue,citecolor=blue,urlcolor=black}

\usepackage{units}

\usepackage{hyperref}
\hypersetup{%
colorlinks,
linkcolor={black},
citecolor={black},
urlcolor={black}}

\begin{document}

\title{Millisecond-lived circular Rydberg atoms in a room-temperature experiment}

 \author{H. Wu}
\affiliation{Laboratoire Kastler Brossel, Coll\`ege de France,
 CNRS, ENS-Universit\'e PSL,
 Sorbonne Universit\'e, \\11, place Marcelin Berthelot, 75005 Paris, France}
 
  \author{R. Richaud}
\affiliation{Laboratoire Kastler Brossel, Coll\`ege de France,
 CNRS, ENS-Universit\'e PSL,
 Sorbonne Universit\'e, \\11, place Marcelin Berthelot, 75005 Paris, France}
 
 \author{J.-M. Raimond}
\affiliation{Laboratoire Kastler Brossel, Coll\`ege de France,
 CNRS, ENS-Universit\'e PSL,
 Sorbonne Universit\'e, \\11, place Marcelin Berthelot, 75005 Paris, France}

\author{M. Brune}
\affiliation{Laboratoire Kastler Brossel, Coll\`ege de France,
 CNRS, ENS-Universit\'e PSL,
 Sorbonne Universit\'e, \\11, place Marcelin Berthelot, 75005 Paris, France}

 \author{S. Gleyzes}
 \affiliation{Laboratoire Kastler Brossel, Coll\`ege de France,
 CNRS, ENS-Universit\'e PSL,
 Sorbonne Universit\'e, \\11, place Marcelin Berthelot, 75005 Paris, France}

\hyphenation{Ryd-berg sen-sing ma-ni-fold}

\begin{abstract}
Circular Rydberg states are ideal tools for quantum technologies, with huge mutual interactions and extremely long lifetimes in the tens of milliseconds range, two orders of magnitude larger than those of laser-accessible Rydberg states. However, such lifetimes are observed only at zero temperature. At room temperature, blackbody-radiation-induced transfers annihilate this essential asset of circular states, which have thus been used mostly so far in specific, complex cryogenic experiments. We demonstrate here, on a laser-cooled atomic sample, a circular state lifetime of more than one millisecond at room temperature for a principal quantum number 60. The inhibition structure is a simple plane-parallel capacitor that efficiently inhibits the blackbody-radiation-induced transfers. One of the capacitor electrodes is fully transparent and provides complete optical access to the atoms, an essential feature for applications. This experiment paves the way to a wide use of circular Rydberg atoms for quantum metrology and quantum simulation.
\end{abstract}

\date{\today}

\maketitle

Rydberg atoms~\cite{GallagherRydbergAtoms1994} are nearly ideal tools for quantum technologies \cite{Saffman_quantum_2010,adams_rydberg_2020}, due to their exquisite sensitivity to external fields and to their enormous mutual dipole-dipole interactions~\cite{LukinDipoleblockadequantum2001,Browaeys_Many-body_2020}. They are extensively used for the development of quantum-enabled sensors~\cite{Sedlacek_microwave_2012,Fan_atom_2015,Faconsensitiveelectrometerbased2016,Downes_full-field_2020}. They lead to spectacular developments for the quantum simulation of spin arrays~\cite{SchollQuantumsimulation2D2021,EbadiQuantumphasesmatter2021}. Most of those experiments are based on ``ordinary'' low-angular-momentum Rydberg states, with a principal quantum number $n$ of a few tens. They are easily accessible from the ground state by laser excitation. Even in a zero-temperature environment, these levels decay relatively rapidly through the emission of optical photons. This is an annoying limitation for some experiments, from metrology to quantum simulations involving large atom numbers.

The circular Rydberg states, with maximum orbital, $\ell$, and magnetic, $m$, quantum numbers ($\ell=m=n-1$), have a much longer lifetime~\cite{RaimondManipulatingquantumentanglement2001}. The $nC$ circular state only decays, in a zero-temperature environment, by the emission of a millimeter-wave photon on the transition towards $(n-1)C$ ($\lambda = 9.6$~mm for $n=60$). This photon is $\sigma^+$--polarized with respect to the quantization axis perpendicular to the electron orbit plane. This single low-frequency decay channel makes the spontaneous emission lifetime two orders of magnitude longer than that of ordinary Rydberg levels ($\sim$ 71~ms for $60C$ and $\sim 500\ \mu$s for $60P$ at zero temperature).

Unfortunately, at room temperature, hundreds of photons per mode in the millimeter-wave Black Body Radiation (BBR) stimulate upward and downward transitions out of the circular states. They reduce drastically their lifetime, which becomes similar to that of low-angular momentum states ($\sim$ 170 $\mu$s for $60C$ and $\sim$~140 $\mu$s for $60P$ at 300~K). So far, this limitation has mostly confined circular Rydberg atoms to specific, quite complex cryogenic experiments~\cite{HarocheExploringquantumatoms2006}.

However, BBR transfers (and even spontaneous emission) can be efficiently inhibited by placing the atoms in a conducting structure below cut-off for most of the relevant frequencies and polarizations~\cite{KleppnerInhibitedSpontaneousEmission1981,VaidyanathanInhibitedAbsorptionBlackbody1981,HuletInhibitedspontaneousemission1985}.
Spontaneous emission inhibition has been observed on a wide variety of systems since the mid-eighties~\cite{GabrielseObservationinhibitedspontaneous1985,HeinzenEnhancedInhibitedVisible1987,JheSuppressionspontaneousemission1987a,BjorkModificationSpontaneousEmission1991,HarocheCavityQuantumElectrodynamics1992,TanakaCavityInducedChangesSpontaneous1995,BayerInhibitionEnhancementSpontaneous2001,LodahlControllingdynamicsspontaneous2004,BienfaitControllingspinrelaxation2016}. In the case of circular states, since the main transitions out of $nC$ are in the millimeter-wave domain, a simple millimeter-size plane-parallel capacitor, perpendicular to the quantization axis, with a spacing $d$ lower than half the wavelength of all the $\sigma$--polarized transitions departing from the circular state, is sufficient to shield the atom from the BBR. 
The ability to control their decay channels resulted recently in a renewed interest in circular Rydberg atoms for quantum simulation~\cite{NguyenQuantumSimulationCircular2018,Meinert_Indium_2020,Cohen_Quantum_2021}.

In this Letter, we report the observation of the efficient inhibition of BBR-induced transfers for circular Rydberg atoms with $n\simeq 60$, prepared out of a laser-cooled cloud of Rubidium atoms in an inhibition capacitor made up of Indium-Tin-Oxide (ITO)-coated glass facing a gold-plated mirror. We observe lifetimes up to 1.1~milliseconds, one order of magnitude larger than those in free space, and close to the values achieved in a finite-temperature cryogenic environment~\cite{Cantat-MoltrechtLonglivedcircularRydberg2020}. One of the capacitor plates being optically transparent, it provides an unrestricted optical access to the inhibition region, an essential feature for quantum technologies.

The experimental set-up is sketched in Fig. 1(a). The capacitor is made up of an ITO-coated glass plate facing a gold-plated copper electrode, both orthogonal to the vertical axis $Oz$. A square with lateral size $a=10$~mm on the gold plated electrode defines the inhibition region. The capacitor spacing is $d=4.1(0.2)$ mm. A static electric field along $Oz$ defines the quantization axis for the atom. For an infinite, ideal capacitor with the same spacing, the BBR-induced transfers out of state $nC$ are efficiently inhibited for $n\ge 58$~\cite{Supplementarymaterial}.  
The set-up is contained in a rectangular glass cell, evacuated to $3\times10^{-9}$ mbar, in which a continuously-operated Rubidium dispenser provides a background Rubidium pressure. The $^{85}$Rb 
atoms are continuously cooled in a standard Mirror-MOT using two counter-propagating 45$^o$ beams, sent through the transparent ITO electrode, and reflecting on the gold-plated one, and two counter-propagating horizontal beams sent through the capacitor spacing. The MOT cloud has a size of about 1~mm.

We excite atoms from the MOT into the $58C$ circular state 
(for details, see~\cite{Supplementarymaterial}).  The scheme begins with a pulsed, stepwise laser excitation towards $58f, m=2$ in the presence of an electric field $\mathbf F$ along $Oz$. The excitation lasers [Fig. 1(a)] define inside the MOT a  $250\times450\times250\ \mu$m$^3$ excitation volume. The `circularization' procedure, thoroughly described elsewhere~\cite{SignolesCoherentTransferLowAngularMomentum2017}, is based on a rapid adiabatic passage between levels of the Stark manifold induced by a $\sigma^+$--polarized radio-frequency (rf) field generated by four electrodes on the sides of the inhibition capacitor [Fig. 1(a)]. It transfers the $58f, m=2$ atoms into $58C$.  Note that dc voltages also applied on the rf electrodes are set to reduce residual  stray field gradients.
From the $58C$ level, we transfer the atom into the circular levels from $54C$ to $60C$ using microwave $\pi$ pulses. We get rid of spurious population in unwanted levels using a combination of rf pulses and partial ionization~\cite{Supplementarymaterial}. 
The end of the preparation process, defining the time origin $t=0$ for the lifetime measurements, occurs at most $89\ \mu$s after the excitation laser pulse.

\begin{figure}
\includegraphics[width=  \linewidth]{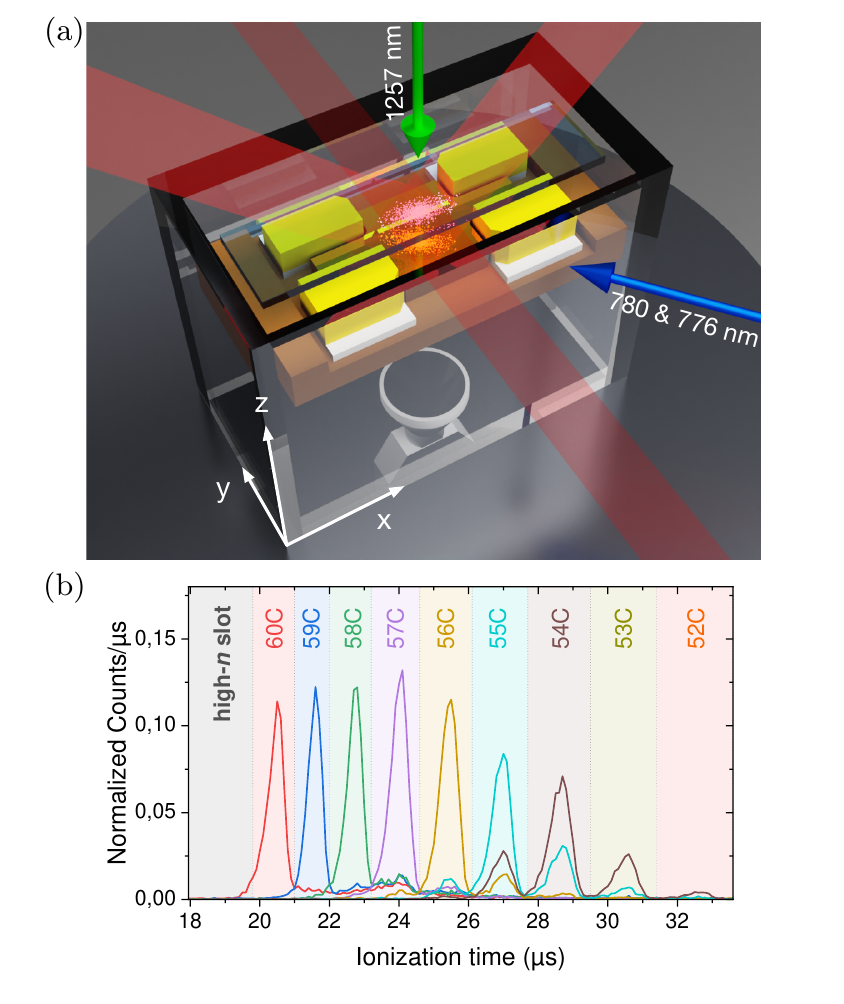}
\caption{ (a) Scheme of the central part of the experiment. The atoms (red particles) are trapped by laser beams (dark red) in a mirror-MOT at the center of a capacitor made up of the transparent, ITO coated glass and of a gold-plated electrode. Additional electrodes on the side (yellow) provide the rf field for circular state preparation.  
The Rydberg excitation lasers are shown as blue and green arrows. The axes are shown in the bottom left. (b) Ionization signals at $\tau=0$ for all the prepared circular states. 
 }
\end{figure}

After a variable delay $\tau$, the atoms are detected. An electric field ionization ramp, starting at $t=\tau$, is applied across the capacitor, reaching at different times the ionization thresholds of different circular states. The resulting ions are sent to a channeltron detector through a 0.2~mm diameter hole, drilled in the bottom plate of the capacitor. We record the number of detected ions as a function of the ionization time, defined as the time delay with respect to the trigger of the ionization ramp.  All relevant levels are ionized before $t=\tau+40\ \mu$s. We detect at most 0.25 ion per preparation sequence. The Rydberg density is thus low enough to preclude any influence of Rydberg-Rydberg interactions.
Fig. 1(b) presents the ionization signals versus ionization time for all prepared circular states at zero delay ($\tau = 0$). The ionization peaks corresponding to different $nC$ states are clearly separated. 
The preparation is not perfect and we observe for all $nC$ states a spurious population in other Rydberg manifolds. From the relative areas of the different peaks, we estimate that the purity of the circular state preparation is between 82\% for $57C$ to 48\% for $54C$.

\begin{figure*}
\includegraphics[width=.85 \linewidth]{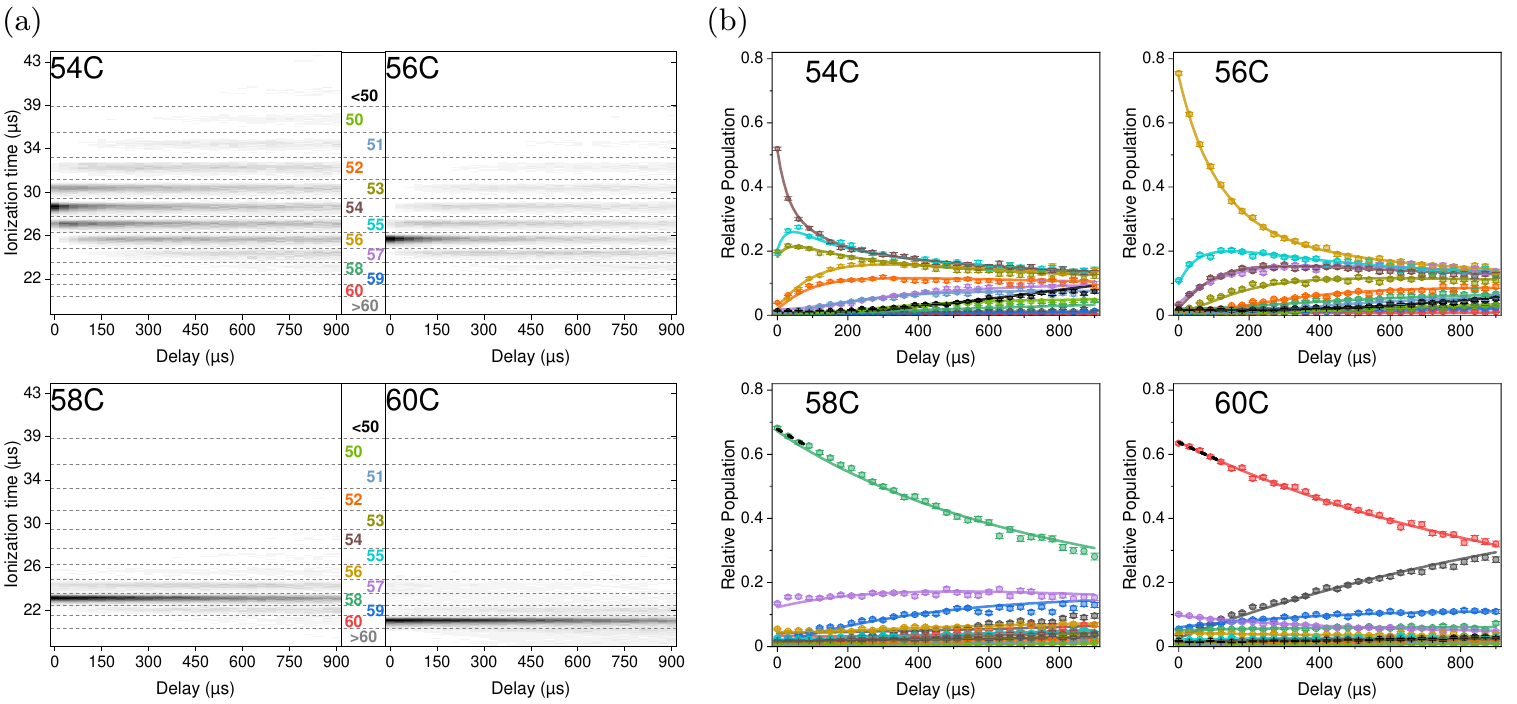} 
\caption{ (a) Grayscale plots of the ionization signal as a function of the delay $\tau$ for different initial levels ($54C$, $56C$, $58C$ and $60C$). For each panel, the grayscale is normalized to the maximum value of the signal at $\tau=0$. Dashed lines represent the time windows used to calculate the population of each manifold. They are labelled by the corresponding value of $n$ between 50 and 60. Levels with $n>60$ or $n<50$ are grouped in two broad detection windows.   
(b) Evolution of the populations of all levels for the same datasets. The color code is given by the color of the labels of the detection windows in (a). The points are experimental, with statistical error bars. Each point corresponds to the detection of  1800 to 9000 ions. The solid lines correspond to the results of the rate equation model fit. The thick black dashed lines show an exponential fit at short times of the prepared state population decay for $n=58$ and $n=60$.}
\end{figure*}

For each initial circular state, $nC$, we record 
ionization signals 
for different delays $\tau$, up to $900\ \mu$s. During this delay, BBR-induced transfers redistribute the Rydberg population among neighboring $n$ manifolds. Simultaneously, other loss mechanisms (atomic sample expansion due to the finite initial Rydberg atoms temperature...) contribute to an overall reduction of the total number of detected ions (by a factor of $\sim$ 3 in $900\ \mu$s). We get rid of these technical losses by renormalizing, for each delay, the total ion count of all detected Rydberg levels to 1. 
Figure 2(a) shows grayscale map plots of the normalized ion signal versus $\tau$ for the initial $nC$ states with $n=54,\ 56,\ 58$ and $60$.  For levels $54C$ and $56C$, we observe a fast redistribution among neighboring manifolds due the strong BBR-induced transitions. For $58C$ and $60C$, this redistribution is much slower, pointing to  a strong inhibition of the BBR-induced transfer rates.

From these data, we get the population of a given manifold by summing the ionization signal over the detection window defined Fig.~2(a). 
Figure 2(b) shows the evolution of the populations of all manifolds for $n$ varying from $50$ to $60$ as a function of time. For $n$'s which are $\geq 61$ or $\leq 49$, we use two broad detection windows. 
The slow evolution of the inhibited levels, $58C$ and $60C$, is conspicuous. From a simple exponential fit to the $60C$ decay at short delays (so that only 10\% of the initial population is lost, making transfers back from neighboring states negligible), we estimate the lifetime of $60C$ to be $1.3$~ms, about $7.6$ times that in free space at room temperature.

In order to get a more quantitative estimate of the inhibition efficiency, we record similar data as a function of $\tau$ for all initial states between $54C$ and $60C$ and fit them on a simple rate equation model for the BBR-induced transfers between adjacent manifolds. 
An atom in the $nC$ state can only decay into the circular state $(n-1)C$. In contrast, absorption of a thermal photon can bring the atom in $(n+1)C$, but can also transfer the atom in a non-circular (``elliptical'') state which ionizes in similar fields as the circular state of the same manifold. 
Due to the ratio of the matrix elements, these transfers to non-circular states are negligible in the absence of inhibition. As a result, for the initial levels with $n\le 56$, the population dynamics is dominated by the $nC\leftrightarrow (n\pm 1)C$ transfers.
Besides, since the number of thermal photons per mode, $n_{\rm th}$, is high at room temperature, the upward and downward transition rates on the $nC\leftrightarrow (n+1)C$ transition are nearly equal. We thus use their common value as a single fit parameter.
For higher quantum numbers
, the circular to circular transitions are strongly inhibited. Transitions towards elliptical levels are no longer negligible and upward transfer rates become larger than downward rates. We model this asymmetry by using different fit parameters for upward and downward transition rates between manifolds $n$ and $(n+1)$ for $n\ge 57$. 

We also use different upward and downward rates for the transfers into the slots gathering all levels above $60C$ and below $50C$ (these slots include many levels and the return rate from them is clearly different from the departure rate into them). We fit as free parameters the initial population of each circular state for each dataset 
(for details on the fitting procedure, see~\cite{Supplementarymaterial}).
\begin{figure}
  \includegraphics[width= 1 \linewidth]{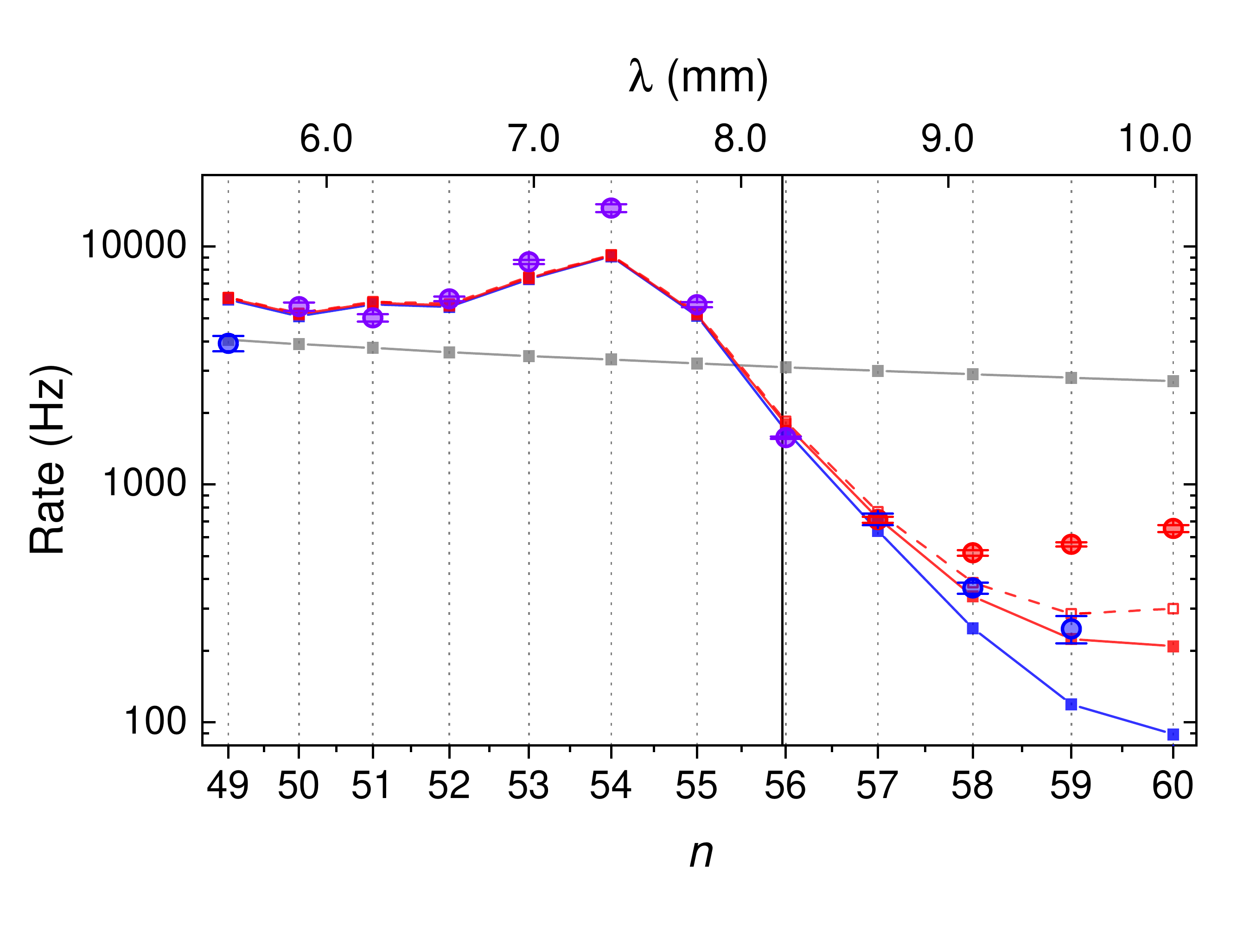}
\caption{Transfer rates between manifolds. The circles are the rates obtained from the fit, with statistical error bars deduced from fits to sub-ensembles of the data (see~\cite{Supplementarymaterial}). For $n\ge 57$, the red and blue circles show the downward $(n+1)\rightarrow n$ and upward $n\rightarrow (n+1)$ rates respectively. The purple circles show the $n\leftrightarrow (n+1)$ rates for $n<57$, when upward and downward rates are assumed to be equal. The transfer rates back from the collective window for high-$n$ or and for low-$n$ are not plotted. The square points are the result of numerical simulations. The blue squares give the $nC\leftrightarrow (n+1)C$ rates, the red squares give the $n\rightarrow (n+1)$ rates when taking into account transfers to the elliptical states. The hollow red squares give the sum of all upwards rate departing from $nC$ (including transfer to the $n+2$ manifold that are neglected in the fit model). The gray squares present the calculated free-space rates. 
 The top scale shows the $nC\leftrightarrow (n+1)C$ transition wavelength $\lambda$. The vertical black line marks the frequency for which $\lambda/2 = d$.}
\end{figure}
The solid lines in Fig.~2(b) present the population evolution resulting from this fit. They are in excellent agreement with the experimental values. 

Figure 3 presents the corresponding transfer rates 
and compares them to the values calculated from a numerical simulation performed with CST Studio~\cite{Supplementarymaterial,CSTweb} 
taking into account the actual electrode structure and assuming a BBR temperature of 300~K. 
The simulated rates are in fair agreement with the fitted ones. Around $\lambda = 7.4$~mm, we observe that the capacitor structure enhances the transition rates as compared to those in free space (black line). In this region, the simulation correctly predicts an enhancement, although it underestimates its effect. This can be explained by an imperfect modelling of the exact resonance of the electrode structure or by a larger microwave field temperature due to the presence of the hot rubidium dispenser inside the UHV chamber.
For $n \ge 56$, the rates are well below the free-space ones, although they are slightly higher than the simulated values. This can be explained by a higher thermal background as well. This might also be due to the finite experiment duration (900$\ \mu$s), which limits the measurement precision for the lowest rate. Nevertheless, the experiment clearly demonstrates thermal transfer inhibition.
For strongly inhibited transitions, the upward rates are higher than the downward rates. This is 
consistent with the fact that upward transitions toward elliptical levels significantly come into play. For the $59\leftrightarrow 60$ transition, the difference between the upward and downward rates is 0.31(3) kHz. It is close to the rate of thermal transfer from 60c toward non circular state predicted by the simulation ($\approx 0.17$~kHz). Our simplified model is thus sufficient to capture the dynamics in the inhibition region.

Finally, we sum the fitted rates  $n\rightarrow (n\pm 1)$ to get the total lifetimes of the $nC$ states (green dots in Fig. 4).  We observe lifetimes up to $1.1 \pm 0.1$~ms for $n=60$.
All the measured lifetimes agree quite well with the simulation results (green line). They are, for the inhibited states, in very good agreement with the estimated lifetime directly based on the exponential fit at short delays (hollow green dots). 
\begin{figure}
\includegraphics[width=1 \linewidth]{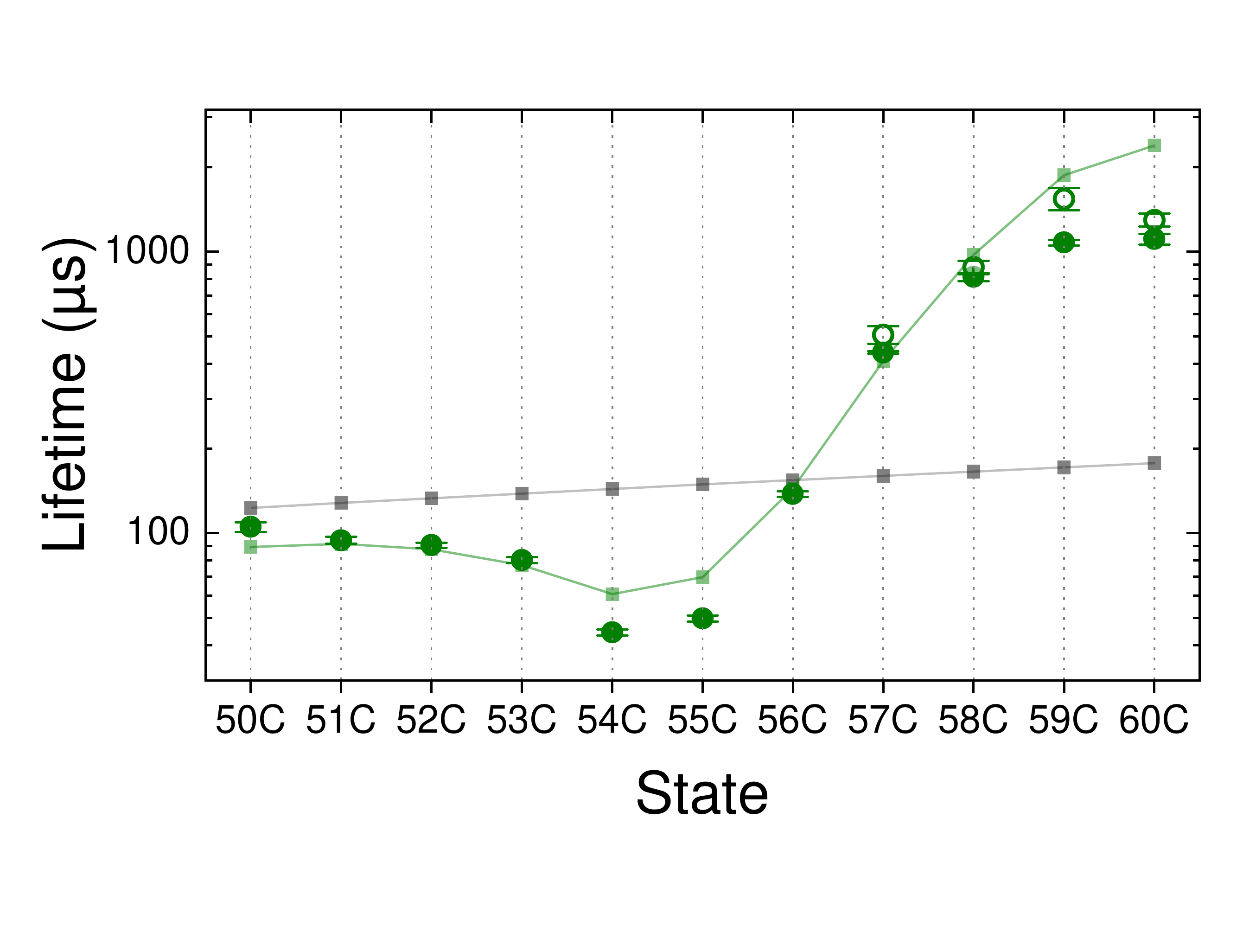}
\caption{Lifetime of the circular states. Solid green circles show the lifetimes deduced from the rates obtained by the fit. The hollow green circles show the results of the exponential fit of the circular state population decays at short delays for large $n$. The error bars are deduced from the fit uncertainties. The square points give the predicted lifetime at 300~K, calculated in free space (black) or inside the electrode structure (green).}
\end{figure}

In conclusion, we have observed lifetime of laser cooled circular states in the millisecond range at room temperature. These lifetimes are one order of magnitude larger than the expected lifetime in free space.  
We have full optical access to the atomic sample, through the transparent electrode of the inhibition capacitor. 
The thermal expansion of the atomic cloud, with a typical millisecond time constant, limits the measurement time, and hence the precision on the longest lifetimes.  Laser-trapped circular atoms~\cite{cortinas_laser_2020}  would remove this limitation. In the inhibition region, the lifetime is limited by the absorption of BBR photons with $\pi$-  or $\sigma$-polarization at high frequency. These decay channels could be inhibited also by placing the atoms in a more complex, 3-dimensional, partly transparent inhibition structure~\cite{CohenQuantumComputingCircular2021}. With these improvements, significantly longer lifetimes are within reach. 
Alternatively, placing the atom in a simple capacitor at zero-temperature environment would realistically lead to lifetimes in the many minutes range~\cite{NguyenQuantumSimulationCircular2018}, with fascinating perspectives for quantum simulation.
The inhibition method demonstrated here would also apply to alkaline earth circular Rydberg atoms~\cite{TeixeiraPreparationLongLivedNonAutoionizing2020,WilsonTrappingAlkalineEarth2022}. The inhibition of BBR transfers thus opens the way for a widespread use of circular atoms in existing and future Rydberg experiments.

\section{Acknowledgements}

We thank Rodrigo Corti\~nas, Brice Ravon and Igor Dotsenko for experimental support, Fernando Gago for fruitful discussions. We acknowledge support from the European Union's Horizon 2020 under grant agreement No 786919 (TRENSCRYBE) and from the Agence Nationale de la Recherche under project 167754 (SNOCAR). H.W. acknowledges a China Scholarship Council No 201806190206.

\section{Supplementary Information}
We describe the preparation of the initial states, the experimental details, the model and CST calculations and the fitting procedures.

\subsection{Preparation of initial states and experimental details}

The circular state preparation starts with a stepwise excitation process using three resonant photons with wavelengths 780~nm ($5S_{1/2}\rightarrow 5P_{3/2}$ transition), 776~nm ($5P_{3/2}\rightarrow 5D_{5/2}$) and 1257~nm ($5D_{5/2}\rightarrow58f$). The three lasers illuminate the MOT cloud simultaneously for 1~$\mu$s at the start of each experimental sequence. During the laser excitation, a small electric field (5~V/m) aligned with the $Oz$ quantization axis, lifts the degeneracy between the $58f$ sublevels with different $|m|$. The frequency and the polarization of the 1257~nm beam are chosen to only excite the $58f$ level with $m=2$.

 After the laser excitation, the field is raised to 195~V/m. The atoms are brought into the $58C$ circular state by an rapid adiabatic passage induced by a $\sigma^+$-polarized rf pulse at 178~MHz, applied while the field is ramped down from 195~V/m to 95~V/m in 6~$\mu$s~\cite{SignolesCoherentTransferLowAngularMomentum2017}. At the end of the `circularization' process, the field is set at 160~V/m.   
 
The purity of the prepared state is not perfect. The circularization is thus followed by a combination of level-selective MW transitions towards the target state, rf transitions within Stark manifolds and partial ionization to increase the purity:
\begin{enumerate}
\item[(i)] For the circular states $nC$ with $n=57-54$, the initial $58C$ level is transferred to the target state by one or two MW $\pi$-pulses (duration $\approx 1 \ \mu$s) selectively addressing the circular-to-circular transitions. The atoms left in Rydberg manifolds with $n' > n$  are then removed by a partial field-ionization ramp that does not affect the target states (circular states remain circular up to their ionization limit).

\item[(ii)]  For $60C$, a two-photon $\pi$-MW pulse first transfers $58C$ to $60C$. The spurious non-circular high-$m$ levels left in the 58 manifold are then driven by a 1~$\mu$s rf pulse towards low-$m$ states in the 58 manifold. This rf pulse is resonant with the transitions between Stark levels within the 58 manifold but not resonant with the ones in the 60 manifold. It thus does not perturb $60C$. Afterwards, the low-$m$ atoms of 58 manifold, which have a lower ionization threshold than $60C$, are removed by partial ionization.

\item[(iii)]  For $58C$, in order to have a higher purity than that provided directly by the circularization processs, we prepare $60C$ as explained in (ii) and transfer the atoms back to $58C$ with a two-photon MW $\pi$-pulse. A partial ionization ramp clears all remaining atoms with $n'>58$. 

\item[(iv)]  Finally, the MW source limitations prevent us from preparing $59C$ from $60C$. We thus prepare the $59C$ state  by a single $\pi$-pulse from the purified $58C$ prepared as in (iii). 
\end{enumerate}

The MOT lasers are always present during the measurement sequence. They do not contribute to the measured lifetime, because circular states have no optical transitions. Photoionization of circular states in these weak laser powers is expected to be negligible~\cite{NguyenQuantumSimulationCircular2018}. The gradient of the MOT magnetic field is estimated to be on the order of 10~Gauss/cm. The induced Zeeman effect is always much lower that the Stark effect.

Dipole-Dipole Rydberg-Rydberg interactions might be an important perturbation, by inducing mixing with non-circular Rydberg states. To rule out such an effect, we reduced the number of atoms initially excited in a Rydberg state by a factor of 5. We did not observe any change in the measured lifetime within statistical errors. 

The compact design of our setup requires the Rb dispenser to be placed very close to the inhibition capacitor. Its operating temperature (700 K) probably makes the effective BBR temperature higher than 300 K. This effect might contribute to the discrepancy between the calculated and observed transition rates.

\subsection{Model and CST calculations}

The modification of the transition rates within the inhibition capacitor is entirely described by the altered classical electromagnetic field-mode density between the plates~\cite{HarocheCavityQuantumElectrodynamics1992, HindsCavityQuantumElectrodynamics1990}. For a numerical estimation of  this classical effect, we compute the power emitted by a classical dipole inside a detailed model of the capacitor, using the CST Studio software suite. The 3D model of the capacitor structure is shown in Fig. 5 (a). The yellow base is the capacitor plate. It is electrically isolated from the 4 rf electrodes above it (also in yellow). All these elements, gold-plated in the set-up, are assumed to have the conductivity of gold at 300 K. The transparent top plate is lead glass material. We mimic the effect of ITO coating with a 2D ohmic sheet covering the whole bottom of this plate with a direct current (dc) sheet resistance of $8 \ \Omega/\text{square}$, matching the value directly measured on the ITO coating by a 4-terminal dc method (the ITO impedance at MW frequency is not expected to be drastically different from its dc value~\cite{Meinert_Indium_2020}).

We compute the power radiated by a dipole antenna with a length on the order of the size of a Rydberg atom and a $\sigma$ or $\pi$-polarization placed at the position of the atoms [marked by a red cone in figure 5(a)]. For each frequency, we compare the total power radiated to the power radiated in free space. The ratio of these quantities directly gives the factor modifying the transition rates at that frequency. The results of this calculation over the frequency range of interest is shown as solid lines in Fig. 5(b). The black and red curves correspond respectively to the $\sigma$ and $\pi$ polarizations. For reference the rate modification factors for an ideal, infinite plane-parallel capacitor with the same 4.1~mm spacing is also represented (dashed lines). The theoretical rates shown in Fig. 3 are then computed using the calculated free-space rates of the relevant transitions at a 300~K background temperature.

\begin{figure}
\includegraphics[width=1\linewidth]{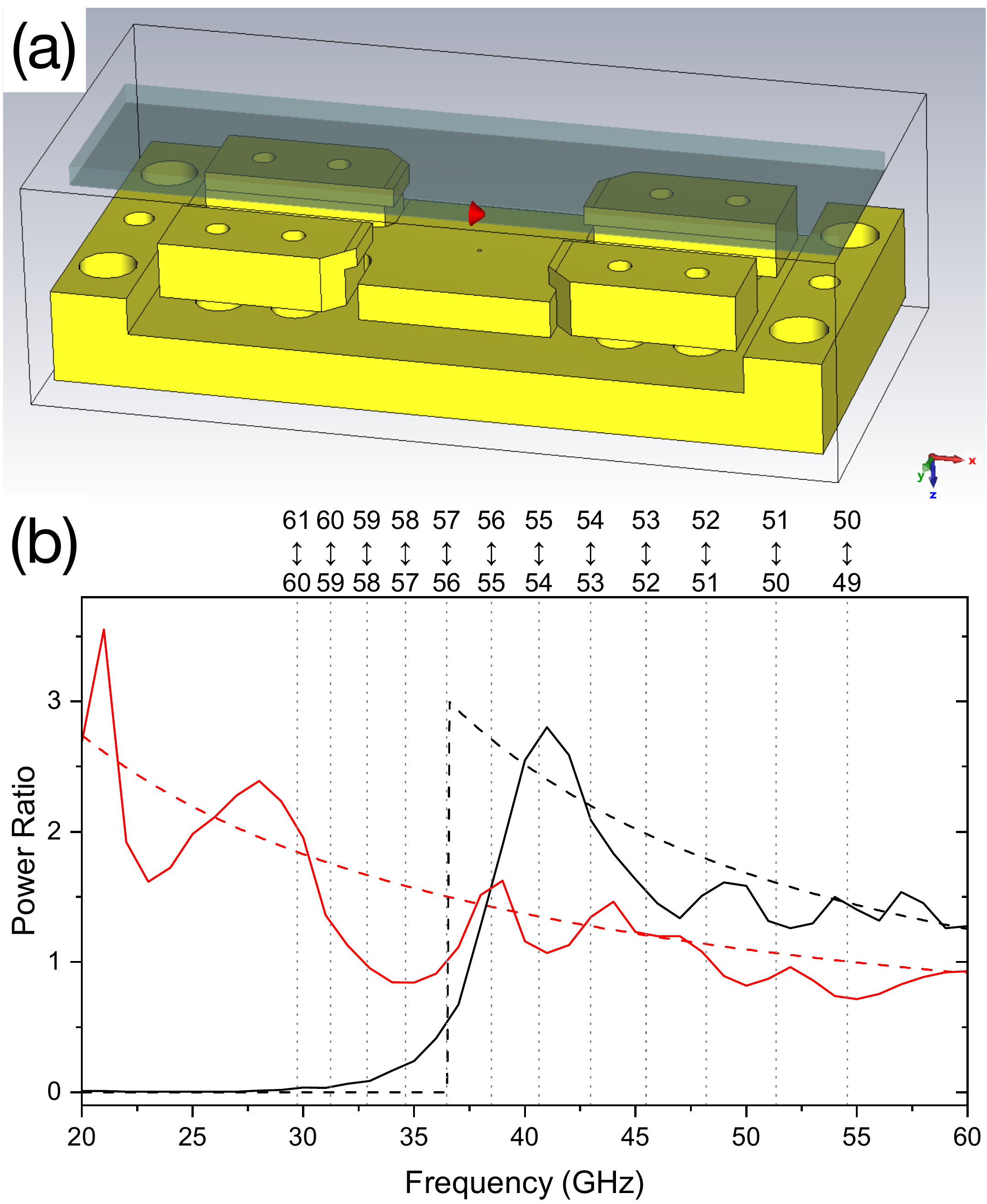}
\caption{ (a) 3D model of the experimental capacitor used for microwave simulations with the CST Studio softwares suite. The capacitor (4.1~mm spacing) is made up of the gold square electrode and of the resistive coating on the glass plate.  Note that the surrounding glass cell is not included in the computation. (b) Ratio of the powers radiated by a classical dipole in the capacitor and in free space. Solid lines correspond to the model capacitor, dashed lines to an infinite ideal plane-parallel capacitor with 4.1~mm plates spacing. Black (red) corresponds to $\sigma$ ($\pi$) radiation. Dotted vertical lines mark circular-to-circular transition frequencies involved in experiment.}
\end{figure}

\subsection{Fitting procedures}

To estimate the transition rates between Rydberg levels, we use a \emph{single} fit of a rate equation model over 16 independent datasets [Figure 2(b) shows only 4 out of these 16 datasets]. They correspond to 16 different experiments that differ in either the chosen initial state (among the seven possible ones, $54C$--$60C$) or the delay range (0 to 150, 0 to 300 or 0 to 900~$\mu$s).
Sequences with smaller delay range are required for a better description of the population transfers at short times, particularly for the non-inhibited transitions.

Figure 6 shows the rates taken into account in the model. For strongly inhibited states ($n \ge 57$), we use two independent fit parameters for the $n\rightarrow n+1$ and the $n+1 \rightarrow n$ rates (red and blue arrows). For levels $n=57$ to $53$, there is little inhibition. As explained in the main text, we thus use the same rate for upward and downward transfers (double headed purple arrows). We have numerically checked this assumption in additional fits. Note that already for $n=57$, the fit result for the two independent rates $58\rightarrow 57$ and $57\rightarrow 58$ are very close  ($710\pm39$~Hz and $715\pm20$~Hz respectively).

Finally, we use two independent parameters for the rates $60\leftrightarrow$ ``high-$n$" and $50\leftrightarrow$ ``low-$n$". Transition back from the ``high-$n$" or ``low-$n$" windows do not represent true inter-level transitions but transitions from detection slots gathering many levels. As a result, they are not represented on Fig. 3. 

In addition to the rates, we also use as fit parameters, for each dataset, the initial populations of all relevant levels at zero delay. The fit is finally performed using the Levenberg-Marquardt algorithm of the SciPy library.

\begin{figure}
\includegraphics[width=1\linewidth]{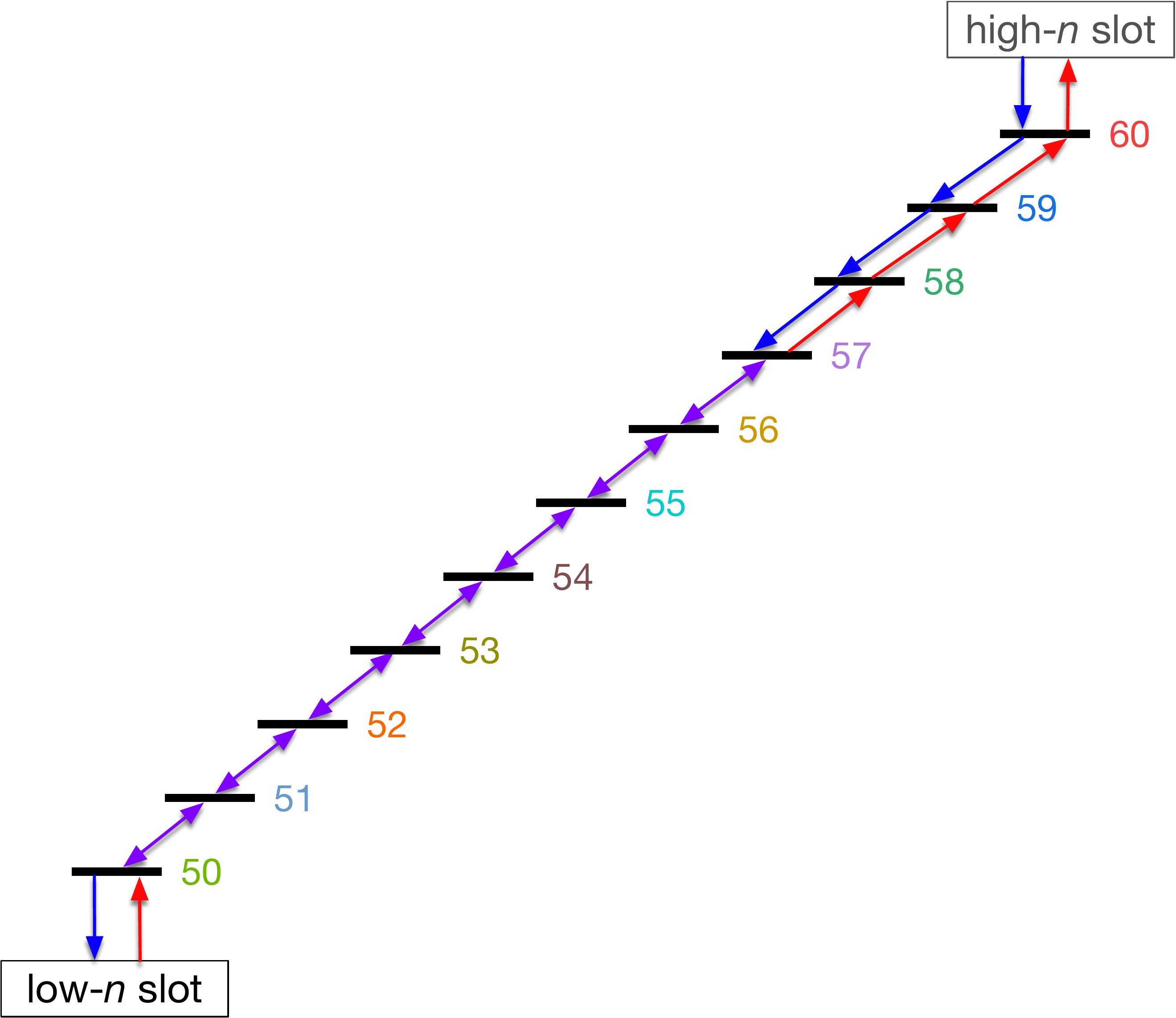}
\caption{Level scheme for the rate equation model used for the fit. Each arrow represents an independent parameter of the fit.}
\end{figure}

In order to estimate the statistical error bars on the rates (shown in Fig. 3), we split each of the 16 datasets in 5 independent sets. We perform separate fits on the resulting 5 subsets. We thus obtain 5 independent sets of rates. The standard errors of these sets provide the error bars in figure 3.

The lifetimes presented in Fig. 4 are calculated from the rates in figure 3. The error bars on these lifetimes are obtained from error propagation of the statistical rates error bars. 

Note that we have performed other fits, with rate equation models based on different transitions schemes, taking into account for instance explicitly sub-sets of the elliptical states in each manifold or direct transitions between $nC$ and the $n+2$ manifolds. The estimated lifetimes always agreed with those of Fig. 4 within the statistical error bars.

\end{document}